\documentclass[%
 reprint,
superscriptaddress,
 amsmath,amssymb,
 aps,
]{revtex4-1}

\usepackage{graphicx}
\usepackage{dcolumn}
\usepackage{bm}
\usepackage[mathlines]{lineno}
\usepackage{hyperref}
\usepackage{xcolor}
\hypersetup{
    colorlinks=true,
    linkcolor=blue,
    citecolor=blue,
    urlcolor=blue,
}
\usepackage{color,soul}

\usepackage[T1]{fontenc}
\usepackage{siunitx}
\DeclareSIUnit\torr{Torr}
\DeclareSIUnit\oersted{Oe}

\begin{document}


\title{Ultra-low current 10 nm spin Hall nano-oscillators}

\author{Nilamani Behera}
\email{nilamani.behera@physics.gu.se}
\affiliation{Physics Department, University of Gothenburg, 412 96 Gothenburg, Sweden.}
\author{Avinash Kumar Chaurasiya}
\affiliation{Physics Department, University of Gothenburg, 412 96 Gothenburg, Sweden.}
\author{Victor H. Gonz\'alez}
\affiliation{Physics Department, University of Gothenburg, 412 96 Gothenburg, Sweden.}
\author{Artem Litvinenko}
\affiliation{Physics Department, University of Gothenburg, 412 96 Gothenburg, Sweden.}
\author{Lakhan Bainsla}
\affiliation{Physics Department, University of Gothenburg, 412 96 Gothenburg, Sweden.}
\author{Akash Kumar}
\affiliation{Physics Department, University of Gothenburg, 412 96 Gothenburg, Sweden.}
\affiliation{Research Institute of Electrical Communication, Tohoku University, 2-1-1 Katahira, Aoba-ku, Sendai, 980-8577, Japan}
\affiliation{Center for Science and Innovation in Spintronics, Tohoku University, 2-1-1 Katahira, Aoba-ku, Sendai, 980-8577, Japan}
\author{Ahmad A. Awad}
\affiliation{Physics Department, University of Gothenburg, 412 96 Gothenburg, Sweden.}
\affiliation{Research Institute of Electrical Communication, Tohoku University, 2-1-1 Katahira, Aoba-ku, Sendai, 980-8577, Japan}
\affiliation{Center for Science and Innovation in Spintronics, Tohoku University, 2-1-1 Katahira, Aoba-ku, Sendai, 980-8577, Japan}
\author{Himanshu Fulara}
\affiliation{Department of Physics, Indian Institute of Technology Roorkee, Roorkee 247667, India}
\author{Johan \AA kerman}
\email{johan.akerman@physics.gu.se}
\affiliation{Physics Department, University of Gothenburg, 412 96 Gothenburg, Sweden.}
\affiliation{Research Institute of Electrical Communication, Tohoku University, 2-1-1 Katahira, Aoba-ku, Sendai, 980-8577, Japan}
\affiliation{Center for Science and Innovation in Spintronics, Tohoku University, 2-1-1 Katahira, Aoba-ku, Sendai, 980-8577, Japan}




\date{\today}

\begin{abstract}
Nano-constriction based spin Hall nano-oscillators (SHNOs) are at the forefront of spintronics research for emerging technological applications such as oscillator-based neuromorphic computing and Ising Machines. 
However, their miniaturization to the sub-50 nm width regime results in poor scaling of the threshold current. Here, we show that current shunting through the Si substrate is the origin of this problem and study how different seed layers can mitigate it. We find that an ultra-thin Al$_{2}$O$_{3}$ seed layer and SiN (200 nm) coated p-Si substrates provide the best improvement, enabling us to scale down the SHNO width to a truly nanoscopic dimension of 10 nm, operating at threshold currents 
below 30 $\mu$A. In addition, the combination of electrical insulation and high thermal conductivity of the Al$_{2}$O$_{3}$ seed will offer the best conditions for large SHNO arrays, avoiding any significant temperature gradients within the array. Our state-of-the-art ultra-low operational current SHNOs hence pave an energy-efficient route to scale oscillator-based computing to large dynamical neural networks of linear chains or two-dimensional arrays.
\end{abstract}

\keywords{spin Hall nano-oscillators, nanoscopic, ultra-low threshold current, micro-Brillouin light scattering}

\pacs{Valid PACS appear here}
\maketitle                         
\section*{Introduction}\label{sec1}
Recent advances in current-induced spin-orbit torques (SOT)\cite{gambardella2011philtrans, sethu2021optimization, sinova2015revmodern} have attracted significant attention due to their energy-efficient applications in memory, communication, spectral analysis, and computation\cite{ramaswamy2016wireless,zhu2018pra, sharma2021electrically,liu2012science, Miron2011nat,Xing2017prappl, Demidov2020jap,Dieny2020natelec, khitun2011non,liu2012prl, demidov2016natcomm, Ranjbar2014, Duan2014b, Zahedinejad2018apl}. Nano-constriction spin Hall nano-oscillators (SHNOs) have emerged as a particularly versatile class of devices, thanks to their straightforward fabrication\cite{Demidov2014apl,mazraati2018apl,Haidar2019natcomm}, direct control using voltage gating\cite{fulara2020natcomm,zahedinejad2022memristive,kumar2022nanoscale} or laser heating\cite{muralidhar2022apl}, and their propensity for mutual synchronization in one\cite{Awad2016natphys,kumar2023robust} and two\cite{Zahedinejad2020natnano} dimensions, which could potentially be used in Ising Machines\cite{albertsson2021ultrafast,houshang2022phase}. In all these applications, power consumption, primarily governed by the SHNO auto-oscillation threshold current, is of utmost importance. While early SHNOs required several mA of current, increased SOT efficiencies\cite{Demasius2016natcomm,utkarsh2021apl,Pai2012,nguyen2015apl}, reduced lateral dimensions\cite{awad2020width,durrenfeld2017nanoscale}, and the addition of perpendicular magnetic anisotropy (PMA)\cite{Fulara2019SciAdv}, have since reduced the threshold current by more than an order of magnitude. The W-Ta/CoFeB/MgO system\cite{behera2022energy, kim2020apl,qian2020spin,sui2017prb, derunova2019giant} is a prime example with its high SOT efficiency, large spin Hall conductivity, and substantial PMA resulting in the first reports of auto-oscillations just below 100 $\mu$A in 50 nm wide SHNOs\cite{behera2022energy}.

In all SHNO demonstrations, the substrate plays a crucial role, where one key property is its thermal conductivity to efficiently dissipate heat generated in the nano-constriction. Although sapphire was initially the substrate of choice in the laboratory, it has since been replaced by CMOS-compatible high-resistance Si (HiR-Si). However, the typical resistivity of 10--40 k$\Omega\cdot$cm is far from insulating, and since the W-Ta/CoFeB/MgO material system is quite resistive, current shunting through the HiR-Si substrate cannot be neglected. This problem rapidly worsens when attempting to scale down the SHNO width below 50 nm.

\begin{figure*}
\centering
\includegraphics[width=14cm]{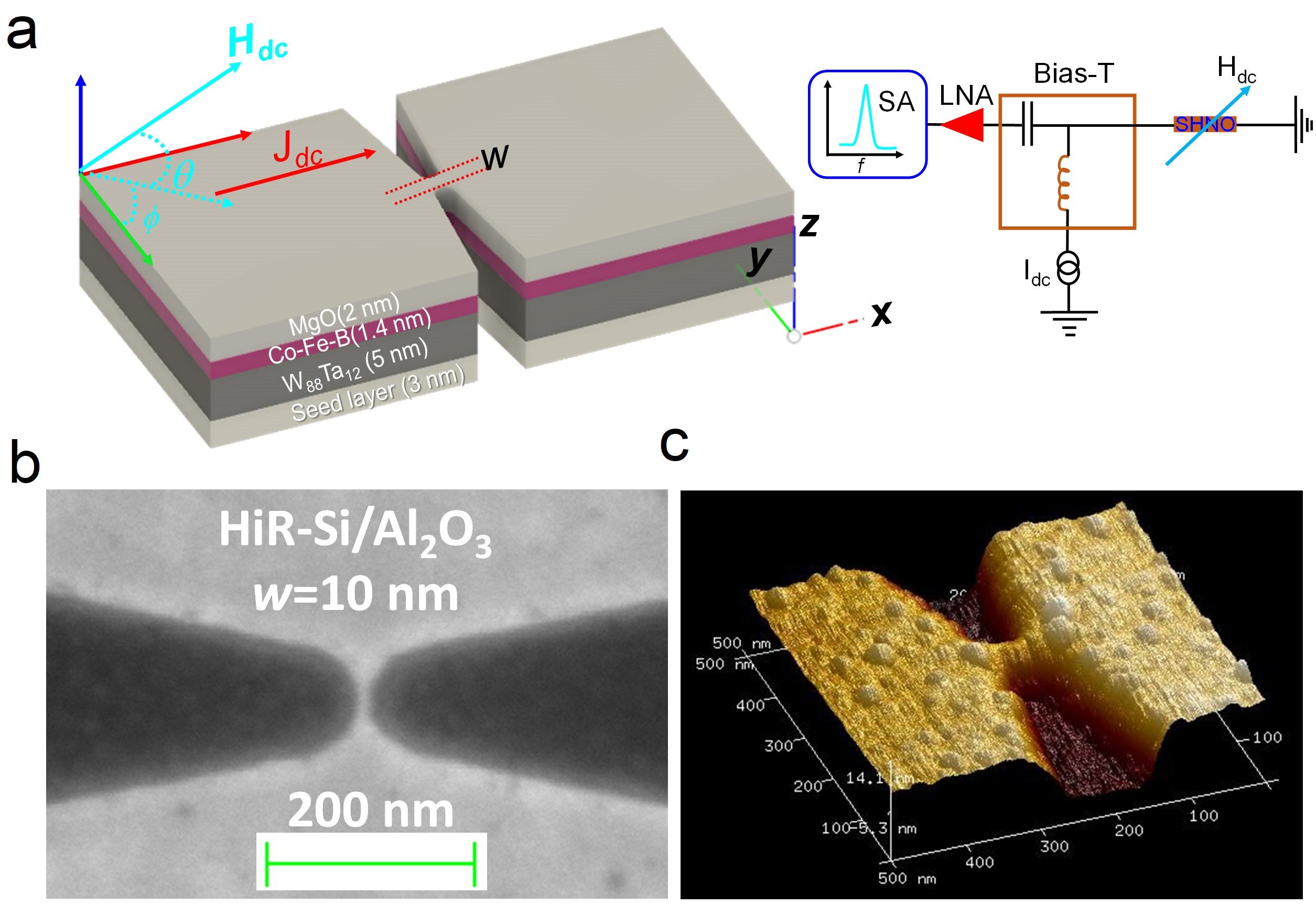}.
\caption{Device description. (a) Schematic of a nano-constriction-based SHNO comprised of a W$_{88}$Ta$_{12}$(5 nm)/Co$_{20}$Fe$_{60}$B$_{20}$(1.4 nm)/MgO(2 nm) stack, and 
a schematic of the measurement setup. (b) Scanning electron micrograph (SEM), and (c) atomic force micrograph (AFM) of a 10 nm wide SHNO grown on an Al${_2}$O${_3}$ seed. 
}
\label{fig:1}
\end{figure*}

Here, we fabricate W-Ta/CoFeB/MgO-based nano-constriction SHNOs over a wide range of widths, $w = 10$--$150$ nm, and study the scaling of their resistance, magnetoresistance, and auto-oscillation threshold current by using both electrical microwave power spectral density (PSD) measurements and optical micro-Brillouin light scattering ($\mu$-BLS) microscopy, while exploring different substrates and seed layers. The main advantage of using $\mu$-BLS microscopy in addition to the electrical measurement is the better sensitivity to detect auto-oscillation signal as well as sub-micron spatial resolution of the measurement. In three of the material stacks, HiR-Si is used as the substrate with \emph{i}) no seed, and seeds of \emph{ii}) 3 nm of Al$_{2}$O$_{3}$, and \emph{iii}) SiO$_{2}$, respectively. A fourth set of SHNOs was deposited on p-Si with 200 nm of SiN. We develop an exact model for the SHNO resistance \emph{vs.}~nano-constriction width and find an excellent agreement within three of the material stacks but significant deviations for the SHNOs fabricated directly on HiR-Si without any insulating seed. This deviation correlates with a width-dependent AMR and very poor scaling of the auto-oscillation threshold current density. In contrast, the use of 3 nm of Al$_{2}$O$_{3}$ 
or SiN (200 nm) coated p-Si substrates results in much better scaling and excellent device yield down to 10 nm, with record low auto-oscillation threshold currents of only 28 and 26 $\mu$A, respectively; this is about a factor of 3--4 better than SHNOs fabricated directly on HiR-Si. Finally, using COMSOL simulations of heat transport in large two-dimensional SHNO arrays, we find that 3 nm of Al$_{2}$O$_{3}$ will be far superior to 200 nm of SiN, thanks to its much higher thermal conductivity, which reduces the temperature gradients.

The demonstration of 10 nm SHNOs operating at 26 $\mu$A has a number of consequences for applications. First, it allows the use of minimal-size transistors providing 26 $\mu$A at a size of 24 nm$^2$ for single SHNOs, relatively lesser area as compared with the 16 nm (48 nm pitch) \cite{Wu2013A1F} and 14 nm (42 nm pitch) \cite{natarajan201414nm,lin2014high} FinFET transistor technology. 
Second, if fabricated in the form of two-dimensional arrays, it should be possible to place 10 nm nano-constrictions at about 24 nm separation, fitting 1600 SHNOs within less than 1 $\mu m^2$. The impedance of such an array would be about 1.35 k$\Omega$, the total current draw 1 mA at threshold, and the total power consumption hence 1.4 mW. Such low power consumption can be easily dissipated via substrate as we show with COMSOL simulation for single and 40x40 array of SHNOs. Third, as the magneto-dipolar coupling strength between SHNOs scales as 1/$d^3$, where  $d$ is the separation between the oscillators, and dominant within the separation range of $\leq$ 100 nm
\cite{slavin2006theory,slavin2009nonlinear}, one would expect such arrays to mutually synchronize very easily, with strongly improved microwave signal properties as the result.

\section*{Results and Discussion}\label{sec2}
Figure~\ref{fig:1}(a) shows schematics of the SHNO and the measurement set-up. 
Figure~\ref{fig:1}(b) shows a top-view SEM image of a 10 nm wide SHNO grown on a 3 nm Al$_{2}$O$_{3}$ seed (additional SEM images are shown in the supplementary Fig. 1). 
From the SEM image, we can conclude that the width of the SHNOs closely matches the nominal value. Figure~\ref{fig:1}(c) shows an atomic force microscopy (AFM) image of a similar 10 nm nano-constriction, confirming 
that no side walls were formed during the fabrication process.\cite{kumar2022nanoscale}  

\begin{figure*}
\centering
\includegraphics[width=16cm]{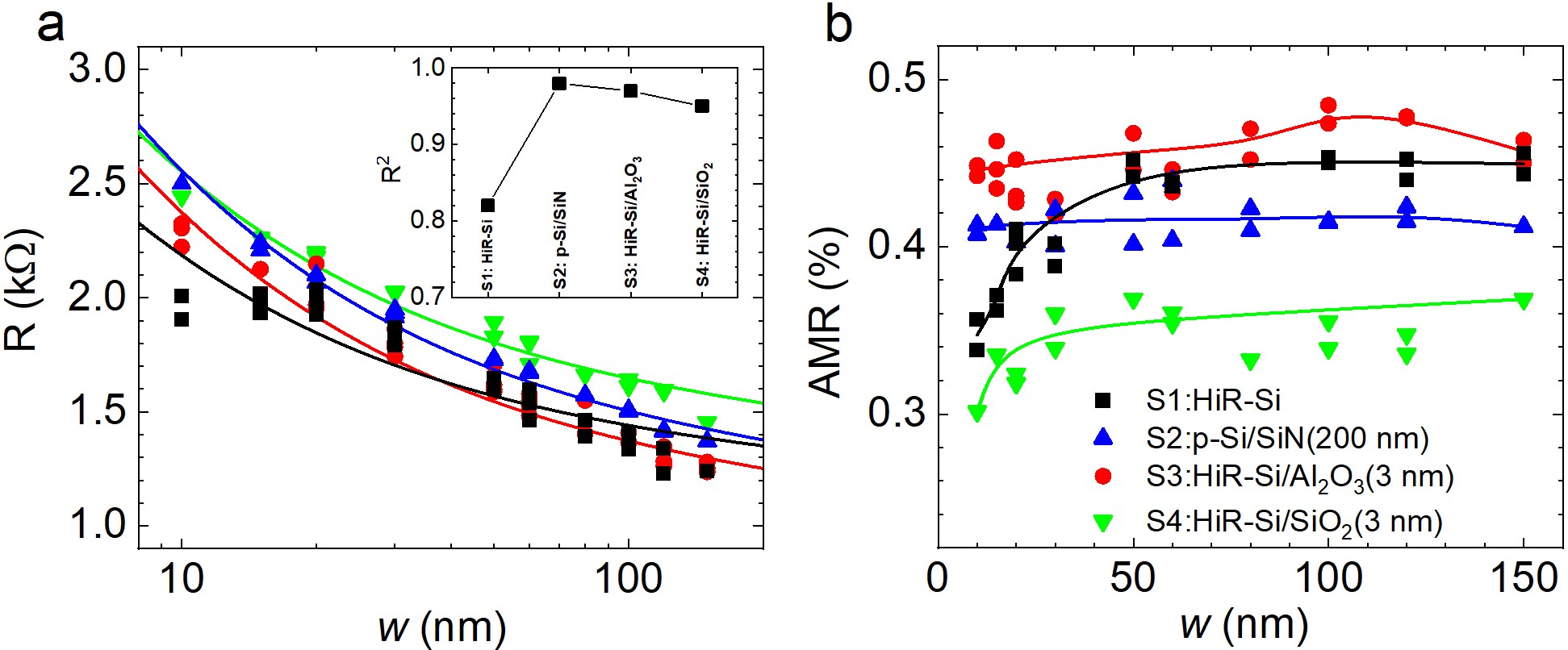}.
\caption{(a) SHNO resistance ($R$) \emph{vs.}~width, and (b) anisotropic magneto-resistance (AMR) \emph{vs.}~width, for all devices. The solid lines in the plot (a) correspond to fits to the model developed in section N2 of the supplementary materials. The inset of Fig. (a) corresponds to best R$^2$ fit value of the theoretical model. The solid lines in plot (b) are guide to the eyes.
}
\label{fig:2}
\end{figure*}

The 
device resistance ($R$) as a function of 
nano-constriction width ($w$) of all SHNOs are plotted on lin-log scales in Fig.~\ref{fig:2}(a) together with fits to a model described in detail in Supplementary Note 2. 
The model takes into account the exact nano-constriction geometry and accurately describes the non-linear $R$ \emph{vs.}~$w$ dependence for the three material stacks with truly insulating substrates/seeds, but less so for the material stack deposited directly onto HiR-Si. The deviation of the HiR-Si data set becomes increasingly severe with decreasing widths, where $R$ saturates around 2 k$\Omega$, consistent with current leakage through the substrate.  
In the inset of Fig.~\ref{fig:2}(a), we quantify the quality of the fit by its coefficient of determination, $\mathcal{R}^2$, and find almost an order of magnitude bigger difference for the HiR-Si data set, clearly separating it from the other data sets.

\begin{table*}[ht]
    \centering
    \caption{Effective magnetization, Gilbert damping, and damping-like SOT efficiency of the material stack grown on different seed layers.}
    \begin{tabular}{|c|c|c|c|}\hline
        Seed layer & $\mu_{0}M_{\mathit{eff}}$ (T) & $\alpha_{\mathit{eff}}$ & $\theta_{\mathit{DL}}$ \\ \hline
        HiR-Si & 0.2492 $\pm$ 0.0011 & 0.0274 $\pm$ 0.0005 & -0.564 $\pm$ 0.004\\
        p-Si/SiN (200 nm) & 0.2533 $\pm$ 0.0011 & 0.0286 $\pm$ 0.0005  & -0.644 $\pm$ 0.004 \\
        HiR-Si/AlO$_x$ (3 nm) & 0.2716 $\pm$ 0.0004 & 0.0255 $\pm$ 0.0005 & -0.647 $\pm$ 0.004\\
        HiR-Si/SiO$_{2}$ (3 nm) & 0.2990 $\pm$ 0.0004 & 0.0367 $\pm$ 0.0007 & -0.641 $\pm$ 0.016 \\ \hline
    \end{tabular}
    
    \label{tab:1}
\end{table*}

Measurements of the anisotropic magnetoresistance (AMR) corroborate these results. Figure~\ref{fig:2}(b) shows a plot of the 
AMR \emph{vs.}~$w$ for all SHNOs. 
Within the scatter of the data, the 
AMR is largely independent of $w$, except in the HiR-Si case, where a clear downwards trend can be observed for $w <$ 50 nm, again consistent with current leakage through the substrate. We note that the absolute AMR of the SHNOs fabricated on SiO$_{2}$ seeds show significantly lower AMR, while those on an Al$_{2}$O$_{3}$ seed show the highest AMR, likely reflecting the intrinsic quality of the material stack. 

To determine the magnetodynamic properties of the different material stacks, we carried out spin torque driven ferromagnetic resonance (ST-FMR) measurements on $6\times12~\mu m^2$ micro-bars (see supplementary Note 5).  
All stacks showed strong, reproducible, and linear current dependencies of the resonance linewidths at different magnetic fields, allowing us to accurately extract the effective magnetization ($M_{\mathit{eff}}$), the damping ($\alpha_{\mathit{eff}}$), and the damping-like spin-orbit torque ($\theta_{\mathit{DL}}$) as seen in Table~\ref{tab:1}. A number of interesting conclusions can be drawn from these data. First, $M_{\mathit{eff}}$ is found to be the lowest for the two stacks grown directly onto their substrates, without any sputtered Al$_{2}$O$_{3}$ or SiO$_{2}$ seed, which indicates that they have slightly stronger perpendicular magnetic anisotropy (PMA); it is noteworthy that the reduction in PMA is significantly worse for SiO$_{2}$ than for Al$_{2}$O$_{3}$. Second, the Gilbert damping is about 50~\% higher for the SiO$_{2}$ seed than for the other three stacks. We hence conclude that the SiO$_{2}$ seed is overall detrimental to the growth of the material stacks as it results in lower AMR, weaker PMA, and much greater damping. Finally, the damping-like SOT is essentially identical for all stacks except for the HiR-Si substrate, which is about 15~\% weaker. This is again consistent with a corresponding current leakage through the HiR-Si substrate. 

\begin{figure*}
\centering
\includegraphics[width=16.5cm]{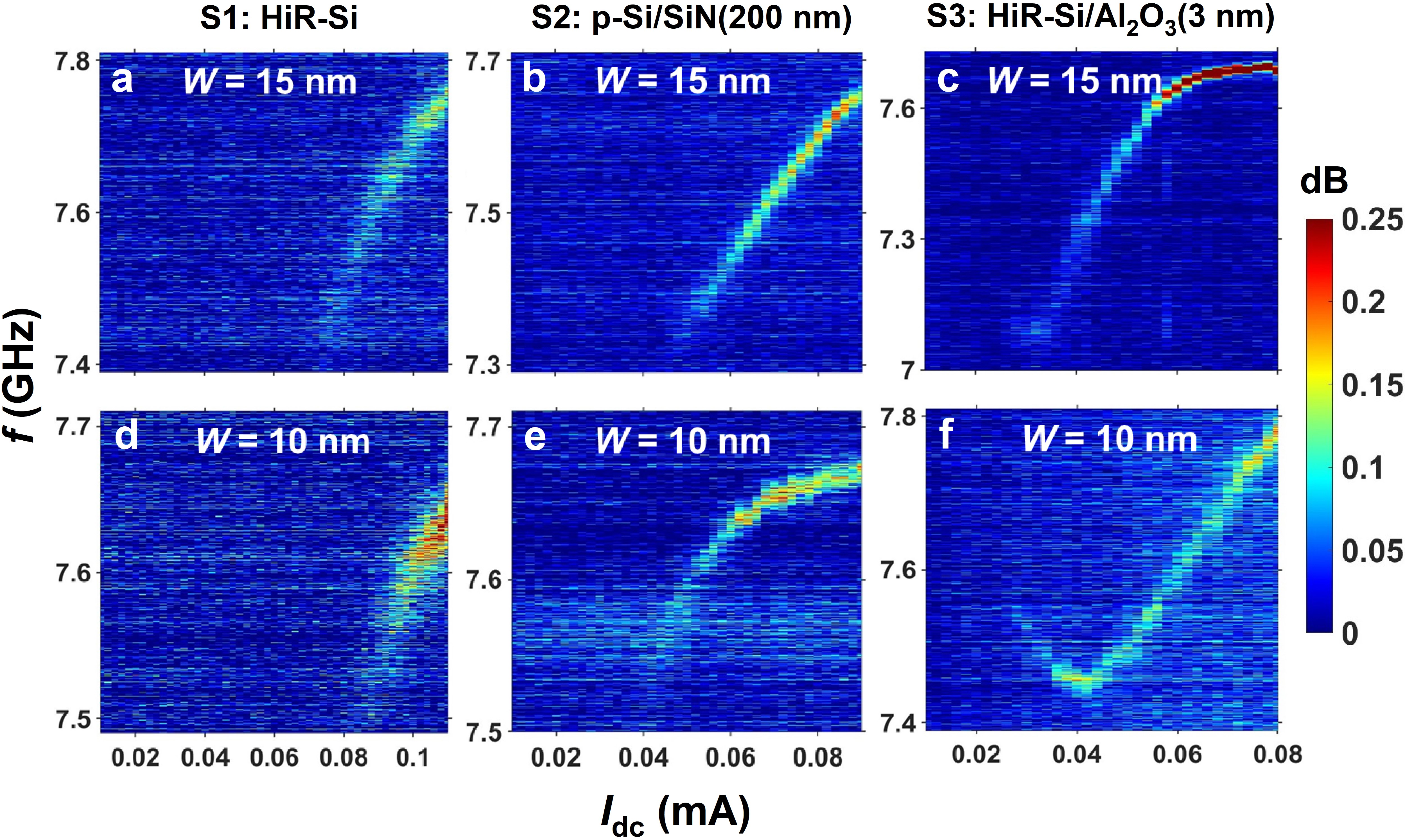}
\caption{ PSD plots of magnetization auto-oscillation versus current for all SHNOs of 15-10 nm nano-constriction width.
The measurements are performed at an in-plane angle of 22$^{\circ}$, and OOP angle of 60$^{\circ}$ and constant magnetic field of 0.3 T.}
\label{fig:3}
\end{figure*}

Next, $I_{dc}$ driven frequency-dependent magnetization auto-oscillation measurements were carried out for all SHNOs of different nano-constriction widths.
The power spectral density (PSD) plots of the smallest SHNOs, $w=$ 15 nm and 10 nm, 
with substrates/seeds S1:HiR-Si, S2:p-Si/SiN(200 nm), and S3:HiR-Si/Al${_2}$O${_3}$(3 nm) are shown in Fig.~\ref{fig:3}. It is noteworthy that we could not observe any microwave voltage signal from the SHNOs based on S4:HiR-Si/SiO${_2}$(3 nm) at these ultranarrow widths, most likely a consequence of their worse magnetoresistive and much worse magnetodynamical properties. In the other SHNOs, we observed a non-monotonic current dependency of auto-oscillation frequency at lower $I_{dc}$ followed by a monotonic behavior at higher $I_{dc}$. This is the fingerprint of an auto-oscillation mode that starts out as localized to one of the nano-constriction edges and then moves inwards to expand across the nano-constriction
\cite{Fulara2019SciAdv, Dvornik2018prappl}. 
For $w=$ 15 nm, 
the auto-oscillation 
threshold currents are S1: 80 $\mu$A, S2: 45 $\mu$A, and S3: 30 $\mu$A 
respectively. These values are nearly one order of magnitude lower compared to earlier works\cite{durrenfeld2017nanoscale,Fulara2019SciAdv,awad2020width,mazraati2018apl}. Clearly, the 
Al${_2}$O${_3}$ seed (S3) results in 
the lowest threshold currents. 
Similarly, for 10 nm wide SHNOs, 
we observed a minimum $I_{th}\sim$~28~$\mu$A for the Al${_2}$O${_3}$ seed (S3) 
as compared to 40~$\mu$A for S2:p-Si/SiN(200 nm) and 100~$\mu$A for S1:HiR-Si, respectively.

\begin{figure*}
\centering
\includegraphics[width=16.5cm]{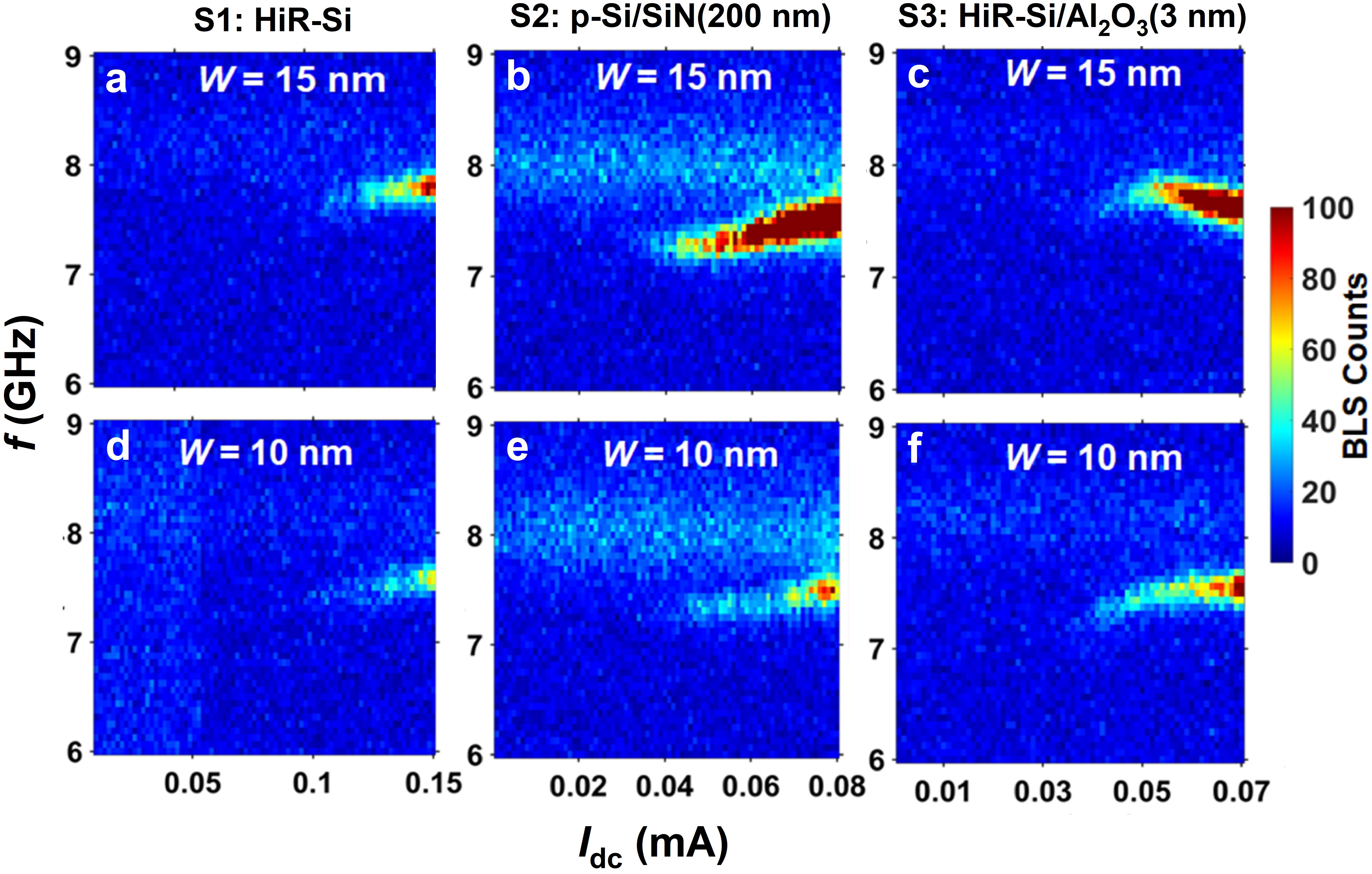}
\caption{$\mu$-BLS spectra of magnetization auto-oscillation versus current for all SHNOs of 15-10 nm nano-constriction widths. The $\mu$-BLS measurements are performed at an in-plane angle of 22$^{\circ}$, and OOP angle of 60$^{\circ}$ and constant magnetic field of 0.3 T. 
}
\label{fig:4}
\end{figure*}

\begin{figure*}
\centering
\includegraphics[width=16cm]{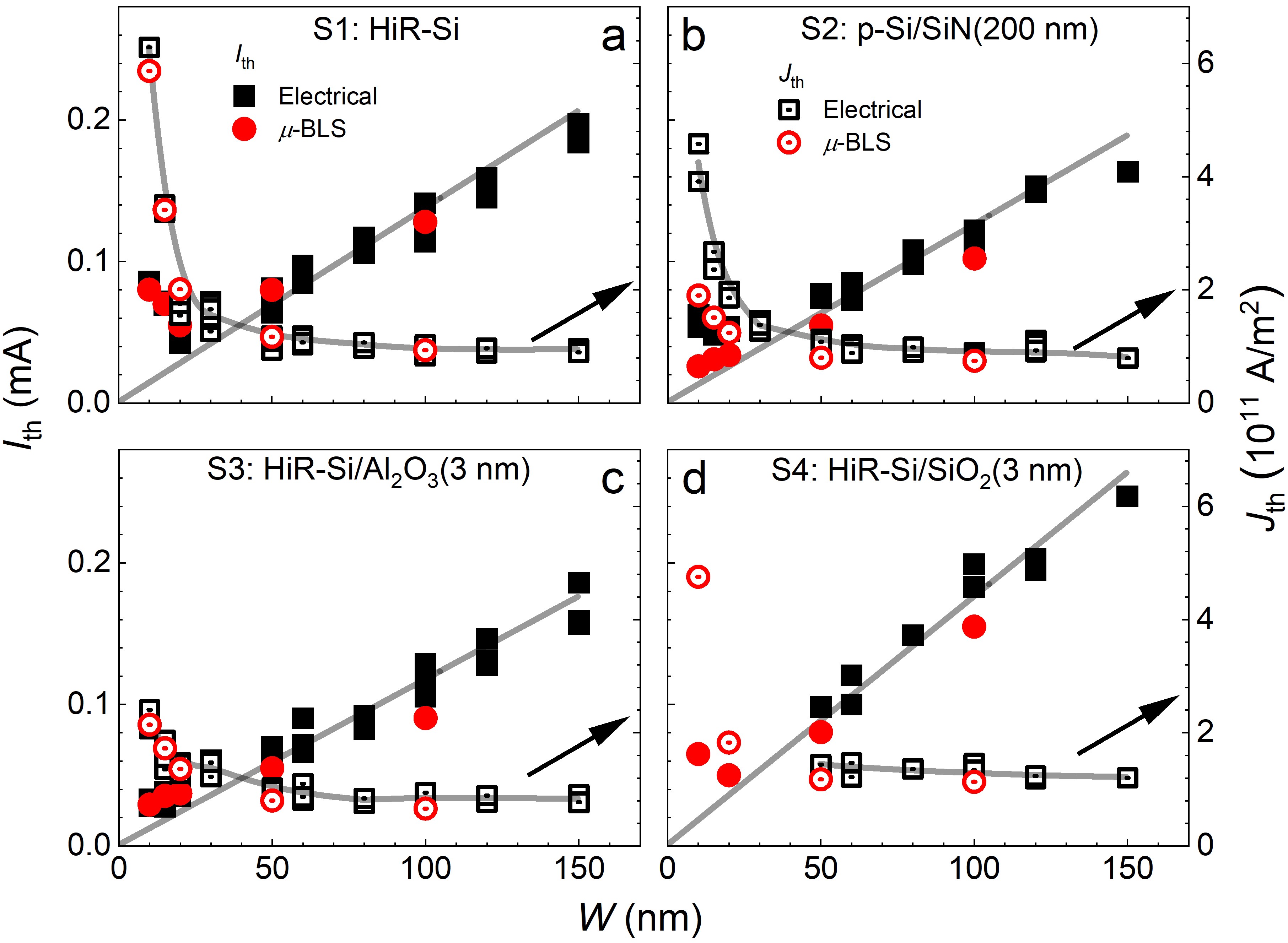}
\caption{Auto-oscillation threshold current (solid symbols; left y-axis) and threshold current density (open symbols; right y-axis) as functions of the nano-constriction widths for all SHNOs. The lines through the solid symbols are linear fits forced through the origin. The lines through the open symbols are guided to the eye.
}
\label{fig:5}
\end{figure*}

The much larger $I_{th}$ 
at smaller $w$ 
of the S1:HiR-Si 
SHNOs is again consistent with parasitic current leakage through the substrate. The leakage can then clearly be mitigated by the insertion of ultrathin insulating seeds. It is interesting to further corroborate this picture by investigating how $I_{th}$ scales with $w$ for the different substrates/seeds. We will first discuss $I_{th}$ extracted from electrical measurements, as measured in Fig.~\ref{fig:3}, and later add complementary optical measurements using micro-Brillouin Light Scattering (BLS) on the same devices in Fig.~\ref{fig:4}. Figure~~\ref{fig:5} summarizes both the electrically and optically determined $I_{th}$ and threshold current densities ($J_{th}$) as a function of $w$ for all four types of material stacks, with filled symbols indicating the $I_{th}$ and hollow symbols the corresponding $J_{th}$.

The electrically determined $I_{th}$ is a linear function of $w$ for $w \geq$ 50 nm. This trend is similar to previous reports\cite{awad2020width,mazraati2018apl}, where the extracted $J_{th}$ is largely independent on $w$ down to 50 nm. However, upon further reduction of $w$ towards 10 nm, $I_{th}$ clearly deviates from the linear behavior and $J_{th}$ rapidly increases. This increase is about 2.5x greater for S1:HiR-Si, compared to S3:HiR-Si/Al${_2}$O${_3}$(3 nm). However, the other two stacks also exhibit a substantial increase in $J_{th}$ of about 2x that of S3. 
While the strong deviation from linear scaling in S1:HiR-Si is consistent with the already observed current leakage in the measurements of R, AMR, and $\theta_{\mathit{DL}}$, the almost as large deviation in $J_{th}$ for S2:p-Si/SiN(200 nm) and S3:HiR-Si/Al${_2}$O${_3}$(3 nm) is harder to reconcile with the earlier conclusion, from the $w$ scaling of their R, AMR, and $\theta_{\mathit{DL}}$, that they do not suffer from any leakage current. As a consequence, we started to question the validity of the electrical measurements, wondering whether the very weak electrical signal might add extrinsic experimental uncertainties when extracting 
$I_{th}$. For example, stacks S2 and S4 both showed substantially weaker AMR and higher $\alpha_{\mathit{eff}}$ compared to S3. As both AMR and $\alpha_{\mathit{eff}}$ directly govern the strength of the auto-oscillation signal, they might affect the apparent $w$ scaling in a detrimental manner. We hence decided to complement the electrical measurements with optical micro-Brillouin Light Scattering ($\mu$-BLS) imaging of the spin wave intensity, allowing us to remove the variation in AMR from the equation and benefit from the higher sensitivity of $\mu$-BLS.  

The details of the $\mu$-BLS setup can be found elsewhere\cite{Sebastian2015MicrofocusedBL}. In the $\mu$-BLS experiment, the magnetic field condition is kept identical to the one used during the electrical measurements described earlier. The $I_{dc}$ induced magnetization auto-oscillation measurements using $\mu$-BLS are carried out for all SHNOs of different constriction widths. The $\mu$-BLS spectra of SHNOs with constriction widths of 15 and 10 nm corresponding to S1, 
S2, 
and S3 
are shown in Fig.~\ref{fig:4}. Here, one can clearly see the auto-oscillating mode and also weak broad traces of the ferromagnetic resonance (FMR) mode around 8 GHz. We define the auto-oscillation threshold as the current at which the BLS intensity starts to grow exponentially. The BLS intensity of the auto-oscillating modes infers a rapid increase of the magnon population with the applied positive current $I_{dc}$ in the vicinity of the threshold value, while the BLS intensity of the FMR remains the same. The observed characteristic behavior of the auto-oscillating mode is consistent with that of the electrical measurements shown in Fig.~\ref{fig:3}. For 15 nm width SHNOs, the auto-oscillation starts at threshold currents of S1: 70~$\mu$A, S2: 31~$\mu$A, and S3: 35~$\mu$A, respectively. Compared to the threshold currents estimated from the electrical measurements, we observed approximately a 10\% and 30\% reduction in $I_{th}$ for S1 
and S2, 
and a 15\% increase in $I_{th}$ for S3, respectively. 
Furthermore, in the case of 10 nm width SHNOs, we observed a minimum $I_{th}$ $\sim$ 26 $\mu$A for S2 
compared to 29 $\mu$A for S3 
and 80 $\mu$A for S1. 
Interestingly, in S4:HiR-Si/SiO${_2}$(3 nm), we were able to detect auto-oscillation signals with $\mu$-BLS for smaller-width SHNOs, where the electrical measurement failed to detect any signal. 
The $\mu$-BLS spectra for 20 nm width SHNOs (S1-S4) are shown in the supplementary Fig. 4.  Finally, both the electrical and $\mu$-BLS experiments validate our claim that the primary reason behind the higher threshold current seen at smaller constriction widths for S1 SHNOs is the effect of leakage current through the high-R Si substrate. For S2-S4 based SHNOs, lower threshold currents are achieved by inserting insulating layers like SiN, Al${_2}$O${_3}$, or SiO${_2}$.

Finally, the low threshold current density enables SHNO operation at low power consumption. It reduces the Joule heating and, as we will see, opens up the possibility of designing very large SHNO arrays needed for neuromorphic computing or for improving the output power and generation linewidth through mutual synchronization. Similar to FinFET nanoscale transistors, one of the main limitations will be the heating \cite{liu2014self}, and heat conduction in very large SHNO arrays differs from that in single SHNOs. In single SHNOs, most of the heat is removed via the metal, and while important, the heat conduction of the substrate is less critical. However, in very large SHNO arrays, there is much less heat conduction through the metal and the substrate and seed become critically important. This can be directly shown using COMSOL simulations of heat transport in single SHNOs and large arrays. 

\begin{figure*}
\centering
\includegraphics[width=17cm]{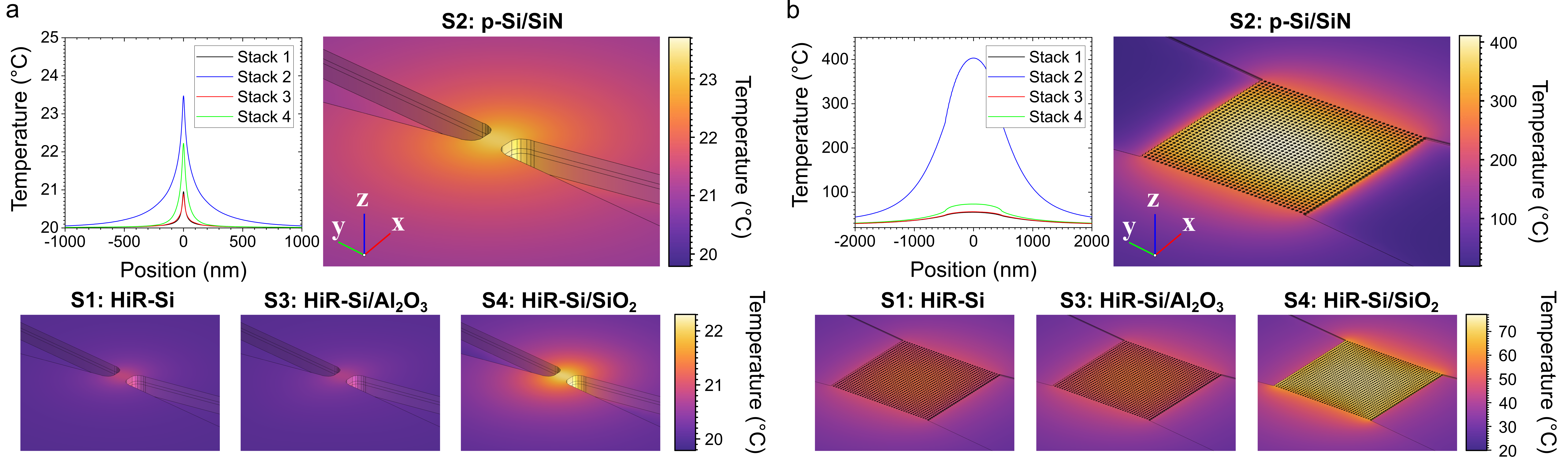}
\caption{ (a) Temperature profiles for the stacks obtained with COMSOL. The profiles are measured along the x-axis. The center of the constriction corresponds to a zero position. The corresponding thermal maps are shown with the captions. (b) Temperature profiles for the stacks obtained with COMSOL. The profiles are measured along the x-axis. The center of the constriction corresponds to a zero position. The corresponding thermal maps are shown with the captions. 
}
\label{fig:6}
\end{figure*}

In order to check the extent of the Joule heating in very large SHNO arrays with the achieved low operational current densities, we performed COMSOL simulations of single SHNOs and arrays of 40x40 SHNOs for the different stacks, S1--S4. Figure~\ref{fig:6}(a) shows the resulting temperature profiles along the x-axis in a single SHNO depending on the stack, for an applied current of 80~$\mu$A. The corresponding thermal maps are shown in the insets. Regardless of the stack, the Joule heating is negligible and results in less than 3~K of temperature increase in the center of the SHNO. Even so, the temperature rise is about 3 times greater with an insulating seed (S3) than with good heat conduction (S1 \& S2). 

The situation is dramatically different in the large arrays. Figure~\ref{fig:6}(b) shows the temperature profiles along the x-axis in 40x40 SHNO arrays, for an applied current of 3.2~mA, \emph{i.e.}~for the same current density as in the single SHNOs. In the case of stack S2, which has a thick layer of SiN with a poor thermal conductivity of 30~W/(m*K), the temperature in the center of the array approaches 400$^\circ$C, which will lead to irreversible damage to the device. However, with good heat conduction through the substrate, the situation is greatly ameliorated, with all arrays operating well below 100$^\circ$C. The combination of the lowest threshold current density and excellent heat conduction in S3:HiR-Si/Al${_2}$O${_3}$(3 nm), hence makes this stack the best choice for any large SHNO array. 

Thus, these novel findings, in addition to an optimized fabrication process, enable us to operate the SHNOs in nanoscopic dimensions as well as with improved energy efficiency. This breakthrough will have significant implications for the development of energy-efficient spintronic devices and applications in various fields, such as telecommunications and data storage.

\section*{Conclusion}\label{sec3}
In summary, we have studied ultra-narrow nano-constriction SHNOs 
based on W$_{88}$Ta$_{12}$(5 nm)/Co$_{20}$Fe$_{60}$B$_{20}$(1.4 nm)/MgO(2 nm) stacks grown on different substrates and seed layers, with particular focus on the scaling of their static, magnetodynamic, and auto-oscillation properties as a function nano-constriction width, $w$. Both electrical microwave power spectral density measurements and optical micro-Brillouin Light Scattering microscopy were used, and found to provide complementary data. For the first time, SHNOs as narrow as 10 nm were experimentally demonstrated, with auto-oscillation threshold currents as low as 26 $\mu$A. Scaling was found to be strongly substrate/seed dependent, in particular for $w \leq$ 50 nm, with the underlying mechanism being current shunting through the HiR-Si substrate. The current shunting issue can be mitigated by using different seed layers, with ultra-thin Al${_2}$O${_3}$(3 nm) showing the best performance, and ultra-thin SiO${_2}$(3 nm) the worst. Whereas a thick SiN(200 nm) seed also worked well in single SHNOs, simulations of large SHNO arrays showed that its much worse heat conduction will lead to temperatures incompatible with CMOS and any prospect for long-term operation. 

\section*{Methods}\label{sec4}
\subsection*{SHNO Fabrication}
The W$_{88}$Ta$_{12}$(5 nm)/Co$_{20}$Fe$_{60}$B$_{20}$(1.4 nm)/MgO(2 nm) optimized stacks are grown on different seed layers and substrates such as high resistive Si (HiR-Si), SiN(200 nm) coated p-Si(100) substrate, Al${_2}$O${_3}$(3 nm) and SiO${_2}$(3 nm) coated HiR-Si substrates (see Fig. ~\ref{fig:1}(a). Hereafter, the list of the sample stacks are used in the present study, S1:HiR-Si (HiR- Si/W$_{88}$Ta$_{12}$(5 nm)/Co$_{20}$Fe$_{60}$B$_{20}$(1.4 nm)/MgO(2 nm)/SiO${_2}$(4 nm)),
S2:p-Si/SiN(200 nm) (p-Si/SiN(200 nm)/W$_{88}$Ta$_{12}$(5 nm)/Co$_{20}$Fe$_{60}$B$_{20}$(1.4 nm)/MgO(2 nm)/Al${_2}$O${_3}$(4 nm)), S3:HiR-Si/Al${_2}$O${_3}$(3 nm)(HiR-Si/Al${_2}$O${_3}$(3 nm)/W$_{88}$Ta$_{12}$(5 nm)/Co$_{20}$Fe$_{60}$B$_{20}$(1.4 nm)/MgO(2 nm)/Al${_2}O{_3}$(4 nm)), and S4:HiR-Si/SiO${_2}$(3 nm)(HiR-Si/SiO${_2}$(3 nm)/W$_{88}$Ta$_{12}$(5 nm)/Co$_{20}$Fe$_{60}$B$_{20}$(1.4 nm)/MgO(2 nm)/SiO${_2}$(4 nm)). The growth environment of each layers in W$_{88}$Ta$_{12}$(5 nm)/Co$_{20}$Fe$_{60}$B$_{20}$(1.4 nm)/MgO(2 nm) stacks are kept similar to our previous work \cite{behera2022energy}. Subsequently, the sample stacks are post-annealed at 300 $^\circ$C for 1 hour to crystallize the Co-Fe-B and MgO at the interface. 

After growth, the sample stacks are fabricated into predefined SHNOs of nano-constriction width varies from 150 nm down to ultra-low of 10 nm using E-beam lithography (EBL). Prior to this, the sample surface is covered with negative electron resist, hydrogen silsesquioxane (HSQ, Product number: XR-1541-002 (2\%)) followed by exposure of EBL (Raith EBPG 5200). Nano-constrictions of various widths are defined in the center of $4\times12~\mu m^2$ mesas. In addition, $6\times12~\mu m^2$ micro-bars are also designed to study spin torque ferromagnetic (STFMR) measurements. Later, Ar-ion beam is used for etching by an Oxford Ionfab 300 Plus etcher to transfer these defined patterns in to sample stack. After that, the negative resist was removed, and ground-signal-ground (GSG) co-planar wave-guide (CPW) patterns are defined by the optical lift-off lithography process. Following this, the CPW is defined 
using sputter-deposited Cu(800 nm)/Pt(20 nm) bilayer.
\subsection*{Scanning electron micro-graph and Atomic force microscopy}
Scanning electron micrograph (SEM) measurements are carried out by using Zeiss Supra 60 VP - EDX 
to confirm the actual lateral dimensions of the measured SHNO devices. Atomic force microscopy (AFM) measurements are performed by using the Dimension Icon Atomic Force Microscope 
to investigate the surface topography of the SHNO devices. 
\subsection*{Anisotropic magneto-resistance}
The in-plane angle dependent an-isotropic magneto-resistance (AMR) measurements on different nano-constriction width-based SHNOs are performed at a constant field strength of 0.1 T by using projected field vector magnet. The in-plane field angle $\phi$ = 0$^{\circ}$ is defined as the current and magnetic field being perpendicular (minimum resistance). The observed overall AMR behavior with respect to the difference of perpendicular and parallel orientations of different nano-constriction widths are plotted on the right panel of Fig.~\ref{fig:2}.

\subsection*{Microwave measurements}
The magnetization auto-oscillation measurements are carried out by using a custom-built probe station with the samples mounted at a fixed in-plane angle on an out-of-plane rotatable sample holder in between pole pieces of the electromagnet. The right side of Fig.~\ref{fig:1}a shows the schematic of the magnetization auto-oscillation measurement setup. A positive dc is fed to the SHNO device through the dc port of the high-frequency bias tee at a fixed out-of-plane magnetic field and the resulting auto-oscillation signal is first amplified by a low-noise amplifier of gain 70 dB and thereafter recorded using a Rohde $\&$ Schwarz (10 Hz to 40 GHz) spectrum analyzer with a low-resolution bandwidth of 300 kHz. 

\subsection*{STFMR measurements}
The spin transfer ferromagnetic resonance (STFMR) measurements are performed on the $6\times12~\mu m^2$ micro-bars (see SI N5 for more details) placed at a fixed in-plane angle with the magnetic field to estimate the spin torque efficiencies by using custom made ferromagnetic resonance set-up (see detail in our previous~\cite{behera2022energy}). First, the microwave signal of various frequencies in the range from 3 to 14 GHz was sent to micro-bars at zero dc through the high-frequency port of bias tee and the simultaneous ST-FMR spectra are recorded in lock-in amplifier through the dc port of the bias tee. Then fix the microwave frequency at 8 GHz and supply dc to micro-bars through the dc port of the bias tee to perform the current modulation measurements.

\subsection*{Micro-Brillouin light scattering (${\mu}$-BLS)measurement}
The magneto-optical measurements of the studied samples were performed using the micro-focused BLS technique~\cite{demokritov2007micro}. A monochromatic continuous wave (CW) laser (wavelength = 532 nm; laser power = 0.8 mW) was focused on the center of the nano-constriction by 100x microscope objective (MO) with a large numerical aperture (NA = 0.75) down to 300 nm diffraction limited spot diameter. The external magnetic field was applied at an in-plane angle of 22$^{\circ}$, and OOP angle of 60$^{\circ}$, and a constant magnetic field magnitude of 0.3 T (which is identical to the one used during the electrical measurements). The inelastically scattered light from the sample was collected by the same MO and analyzed with a Sandercock-type six-pass Tandem Fabry-Perot interferometer TFP-1 (JRS Scientific Instruments). The obtained BLS intensity is proportional to the square of the amplitude of the dynamic magnetization at the position of the laser spot. A stabilization software based on active feedback algorithm (provided by THATec Innovation) was employed in order to get long-term spatial stability during the $\mu$-BLS measurement. 

\subsection*{COMSOL simulation}
We exploited the COMSOL module Electric Currents (ec) to simulate electric currents in the conducting layers and the Heat Transfer in Solids (ht) module to simulate heat dissipation through all the layers. We took into account the 3~nm silicon oxide layer on top of the silicon wafer which has a significantly lower thermal conductivity of 1.4~W/(m*K). The base HiR-Si wafer has a thermal conductivity of 130~W/(m*K). The simulations are performed using the measured resistivity for the thin films i.e. W-Ta (260 $\mu \Omega$-cm) and CoFeB (64 $\mu \Omega$-cm). In order to reduce the simulation time and resources we simulate a limited chip area of 0.5x0.5x0.5~mm. Temperature boundary conditions of 293.15~K are applied at the edges of the simulated area. 

\subsection*{Acknowledgement}
This work was partially supported by the Horizon 2020 research and innovation program No. 835068 "TOPSPIN". This work was also partially supported by the Swedish Research Council (VR Grant No. 2016-05980) and the Knut and Alice Wallenberg Foundation. This work was performed in part at Myfab Chalmers.

\bibliography{ref}


\end{document}